# Spin structure factor and quantum phases of frustrated spin-1/2 chains

Manoranjan Kumar,[1] Aslam Parvej[1] and Z.G. Soos[2]

[1] *S.N. Bose National Centre for Basic Sciences. Block-JD, Sector-III, Kolkata 700098, India.*
[2] *Department of Chemistry, Princeton University, Princeton, New Jersey 08544, USA.*

**Abstract** – The static structure factor $S(q)$ of frustrated spin-1/2 chains with isotropic exchange and a singlet ground state (GS) diverges at wave vector $q_m$ when the GS has quasi-long-range order (QLRO) with periodicity $2\pi/q_m$ but $S(q_m)$ is finite in bond-order-wave (BOW) phases with finite-range spin correlations. Exact diagonalization and density matrix renormalization group (DMRG) calculations of $S(q)$ indicate a decoupled phase with QLRO and $q_m = \pi/2$ in chains with large antiferromagnetic exchange between second neighbors. $S(q_m)$ identifies quantum phase transitions based on GS spin correlations.

Email: manoranjan.kumar@bose.res.in, soos@princeton.edu

## 1. Introduction

The $J_1J_2$ model with isotropic exchange $J_1, J_2 > 0$ between first and second neighbors is the prototypical frustrated spin-1/2 chain with a bond-order-wave (BOW) phase.[1-15] The Hamiltonian with periodic boundary conditions (PBC) and frustration $g = J_2/J_1 > 0$ is

$$H(g) = \sum_p (\vec{s}_p \cdot \vec{s}_{p+1} + g\vec{s}_p \cdot \vec{s}_{p+2})$$
$$= g(H_A + H_B) + \sum_p \vec{s}_p \cdot \vec{s}_{p+1} \quad (1)$$

$H_A$ and $H_B$ are linear Heisenberg antiferromagnets (HAFs) with PBC on sublattices of odd and even-numbered sites, and $H(0)$ is also an HAF. The ground state (GS) of Eq. 1 is a singlet, S=0. The infinite chain has nondegenerate GS at small $g$ that becomes doubly degenerate[8] at $g^* = 0.2411$, the boundary of the BOW phase with broken inversion symmetry at sites and a finite energy gap[2] $E_m(g)$ to the lowest triplet state. The exact GS at the Majumdar-Ghosh point,[1] $g = 1/2$, are the Kekulé diagrams $|K1\rangle$ and $|K2\rangle$ of organic chemistry that correspond to singlet-paired spins on adjacent sites. Recent studies[16-20] have focused on ferromagnetic $J_1 < 0$ as the starting point for modeling oxides with chains of $s = 1/2$ Cu(II) ions.

Bursill et al.[10] studied the static spin structure factor $S(q;g)$ of the $J_1J_2$ model and took the $S(q_m)$ peak as the effective periodicity $2\pi/q_m$. They compared $q_m$ to chains of classical spins, for which the GS energy of Eq. 1 goes as $\cos\chi + g\cos 2\chi$ where $\chi$ is the pitch angle between successive spins. Minimization gives $\cos\chi = -1/4g$, or $\chi = \pi$ for $g < 1/4$ and a continuous decrease to $\chi = \pi/2$ as $g \to \infty$. Quantum effects[10] are pronounced at small $g$, where $q_m = \pi$ persists to $g = 1/2$. The BOW phase extends to arbitrarily large $g$ according to Bursill et al.[10] and the field theories of White and Affleck[11] and Itoi and Qin[13]. We find instead that the BOW phase terminates at $1/g^{**} \sim 0.40$ at the start of a gapless decoupled phase[21,22] with nondegenerate GS. We return in the Discussion to reasons for reexamining the quantum phase diagram at large $g$.

In this paper, the *magnitude* of $S(q_m;g)$ is applied to the quantum phase diagram of frustrated spin chains. $S(q_m;g)$ diverges when the GS has quasi-long-range order (QLRO) at wave vector $q_m$. The HAF at $g = 0$ has QLRO($\pi$) while the BOW phase has finite $S(q_m;g)$ and spin correlations that are just to nearest neighbors at $g = 1/2$. The quantum transition between the QLRO($\pi$) and BOW phases is the largest $g$ at which $S(\pi;g)$ diverges; as shown in Section 3, this agrees with $g^* = 0.2411$ based[8] on the degeneracy, $E_m = E_\sigma$, of the triplet and lowest singlet excitation. HAFs on sublattices at $1/g = 0$ have QLRO($\pi/2$) and divergent $S(\pi/2;\infty)$. The largest $1/g$ at which $S(\pi/2;g)$ diverges marks the transition from the decoupled to the BOW phase. In our analysis, the frustrated BOW phase with finite $S(q;g)$ is intermediate between phases with dominant QLRO($\pi$) at small $g$ and QLRO($\pi/2$) at small $1/g$.

We obtain $S(q;g)$ using exact diagonalization (ED) of finite $J_1J_2$ models, density matrix renormalization group (DMRG) calculations and extrapolation to the infinite chain. The procedure is general for spin chains. Sections 2 and 3 present S(q;g) results and the size dependence of S(q$_m$;g), respectively. In Section 4 we briefly discuss the gapless decoupled phase and specific challenges of solving H(g) at large g.

## 2. Static structure factor S(q)

The static structure factor S(q) of 1D systems with one spin per unit cell is the GS expectation value

$$S(q) = \frac{1}{N}\sum_{p,r}\langle \vec{s}_p \cdot \vec{s}_r\rangle e^{iq(p-r)}$$
$$= \sum_p \langle \vec{s}_1 \cdot \vec{s}_{1+p}\rangle e^{iqp} \quad (2)$$

The wave vectors in the first Brillouin zone are $q = 2\pi m/N$ with $m = 0, \pm 1, \ldots, N/2$. We define spin correlation functions $C(p,g) = \langle \mathbf{s}_1 \cdot \mathbf{s}_{1+p}\rangle$ at frustration g in Eq. 1 and consider S(q;g,4n) with N = 4n spins that ensure integer total spin $S \leq 2n$ and sublattice spin $S_A, S_B \leq n$. The q = 0 component satisfies $S(0;g) = \langle S^2\rangle/4n = 0$ when the GS is a singlet; the sum of C(p;g) over p is zero; summing over q in the Brillouin zone and taking the limit $n \to \infty$ leads to

$$\frac{1}{4n}\sum_q S(q;g,4n) = \frac{3}{4}$$
$$= \frac{1}{\pi}\int_0^\pi dq\, S(q;g) \quad (3)$$

since C(0,g) = 3/4 for s = 1/2.

If the C(p,g) have finite range, S(q;g) is finite and the sum in Eq. 2 becomes constant once the system size exceeds the correlation length. For even N in Eq. 2, the exact GS at g = 1/2 gives

$$S(q;1/2) = 3(1-\cos q)/4 \quad (4)$$

The size dependence is entirely in the discrete q values. A finite energy gap $E_m(g)$ in the BOW phase indicates a localized GS and finite-range spin correlations. S(q;g) is readily found directly for some g in the BOW phase. We defer to Section 3 the numerical problem of the divergence of S(q$_m$;g).

To illustrate, we choose g in Eq. 1 such that 24 spins is close to the infinite chain. The peak is better seen in the zone $0 \leq q < 2\pi$. Open symbols in the upper panel of Fig. 1 are exact S(q;g,24) at discrete q in Eq. 2 for g = 0.40, 0.50, 0.70 and 1.0. Solid lines are S(q;g,48) with continuous q obtained by DMRG for 48 spins. Almost identical S(q;g) are found except at g = 1 where the dashed line refers to 24 spins and continuous q. The peak at q$_m$ = π for g = 0.40 and 0.50 evolves with increasing g to π/2 and 3π/2 (−π/2).

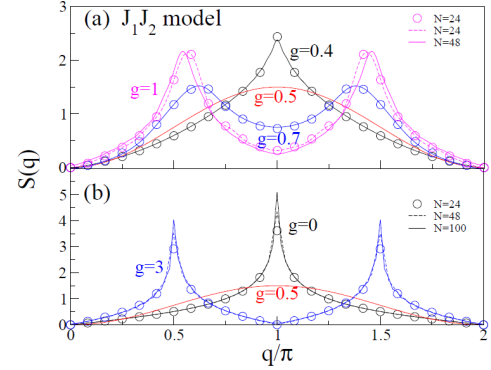

**Fig. 1.** Spin structure factor S(q), Eq. 2, with frustration g in the $J_1J_2$ model, Eq. 1. Open symbols are exact for N = 24 spins and discrete wave vectors q. (a) S(q) is finite in the BOW phase. Solid lines are DMRG results for N = 48 spins and continuous q; the dashed line at g = 1 has 24 spins and continuous q. (b) Solid lines and dashed lines are DMRG results for N = 100 and 48 spins. The S(q) peaks at q = π for g = 0 and at ±π/2 for g = 3 increases with system size in phases with quasi-long-range order.

The HAF is a gapless spin liquid[23] with QLRO(π), algebraically decreasing C(p,0) and divergent S(π;0). At 1/g = 0, we have HAFs on sublattices, QLRO(π/2) and divergent S(π/2;∞). The lower panel of Fig. 1 contrasts S(q;g) at g = 0 and 3 with g = 1/2. Open symbols are exact S(q;g,24) at g = 0 and 3; the dashed and solid lines are DMRG results for 48 and 100 spins, respectively. Quite generally, we have S(q;0,4n) = S(q/2;∞,8n) since both g = 0 and 1/g = 0 correspond to 4n-spin HAFs. The q$_m$ = π peak for 24 spins at g = 0 is almost exactly equal to the q$_m$ = π/2 peak for 48 spins at g = 3. Fig. 1 already suggests that the BOW phase does not extend to g = 3. As shown in Section 3, the lowest-order changes go as

$$S(\pi;g,4n) = S(\pi;0,4n) - A_n g$$
$$S(\pi/2;1/g,8n) = S(\pi/2;0,8n) - B_n/g^2 \quad (5)$$

with $A_n, B_n > 0$. Since the peaks are equal at g = 0 = 1/g, the π/2 peak in finite systems is less sensitive to frustration 1/g << 1 than the π peak is to g << 1.

The wave vector q$_m$ is shown in Fig. 2 as a function of g/(1 + g). Open circles are exact for 24 spins. The peak remains at π up to g = 1/2 and then decreases to $q_m = \pi/2$. Classical spins have pitch angle $q_m$, with $\cos q_m = -1/4g$ for $g > 1/4$, doubly degenerate GS with long-range Néel order up to $g = 1/4$, and a spiral GS with LRO($q_m$) for $g > 1/4$. We find strong quantum effects at large $g$ that compress the BOW phase and lock in $q_m = \pi/2$.

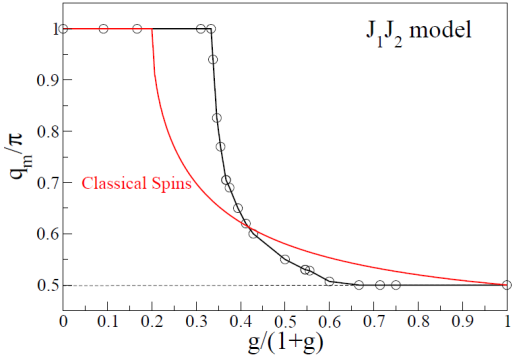

Fig. 2. Wave vector $q_m$ of the structure factor peak $S(q_m)$ of the $J_1J_2$ model with $N = 24$ spins as function of frustration $g/(1 + g)$. The chain of classical spins has pitch angel $\chi = \pi$ for $g \leq 1/4$ and $\chi = \cos^{-1}(-1/4g)$ for $g > 1/4$.

## 3. Phase transitions

The static structure factor identifies the three quantum phases of the $J_1J_2$ model. The location of phase transitions is more demanding. Finite N in Eq. 2 clearly gives finite $S(q)$. We must infer whether $S(\pi;g,4n)$ or $S(\pi/2;g,4n)$ diverges with increasing system size rather than merely becoming large. The numerical problem is to compute all spin correlations $C(p,g)$ in systems of $N = 4n$ spins. We use ED up to 24 spins and a finite DMRG algorithm for larger systems with four spins added per step[15] and cyclic boundary conditions.[24] The algorithm is more accurate than conventional DMRG because adding four spins per step ensures that the sublattices always have $S_A = S_B = 0$ at $1/g = 0$ rather than $S_A = S_B = 1/2$ at every other step. Truncation errors in the sum of the eigenvalues of the density matrix are less than $10^{-10}$ in the worst case when $m = 200$ eigenvalues are kept. Finite size effects increase at large g. DMRG returns $C(p,g)$ whose accuracy can be tested rigorously by comparison to the exact result, $S(0;g,4n) = 0$. We find $S(0;g,100) < 10^{-3}$ in the QLRO($\pi$) phase up to $4n = 100$ and comparable accuracy to $4n = 64$ in the QLRO($\pi/2$) phase.

We also rely on HAF spin correlation functions[23] that establish the divergence of $S(\pi;0)$ or $S(\pi/2;\infty)$. The $q = \pi$ term of Eq. 2 for $4n$ spins is

$$S(\pi;g,4n) = \frac{3}{4} + C(2n,g) + 2\sum_{p=1}^{2n-1} C(p,g)(-1)^p \quad (6)$$

Since $C(p,0)$ goes as $(-1)^p$, the sum is over $|C(p,0)|$. As shown in the inset to Fig. 3, $S(\pi;g,4n)$ is a linear function at small g with slope $-A_n$ and $A_6 = 1.6$ for 24 spins. Finite $g > 0$ is frustrating while $g < 0$ enhances short-range $q = \pi$ order.

Incremental increases of $S(\pi;g,4n)$ from $4n$ to $4n + 4$ spins are shown in Fig. 3 as a function of $100/N$ with $N = 4n + 2$, followed by linear extrapolation to the infinite chain. $S(\pi;0.40,4n)$ converges rapidly as noted in Fig. 1. Within our accuracy, $S(\pi;g)$ diverges at $g = 0.20$ and converges at $g = 0.25$. The estimated $g^*$ between 0.20 and 0.25 based on the structure factor is consistent with, but much less precise than $g^* = 0.2411$ based[8] on $E_m = E_\sigma$. The two methods are independent since the GS determines $S(\pi;g)$ but does not enter in the excited-state degeneracy.

Only spin correlations within one sublattice contribute to $S(\pi/2)$

$$S(\pi/2;g,4n) = \frac{3}{4} + C(2n,g)(-1)^n$$
$$+ 2\sum_{p=1}^{n-1} C(p,g)\cos(\pi p/2) \quad (7)$$

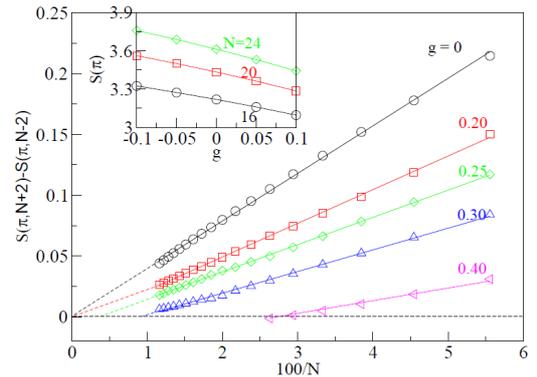

Fig. 3. Incremental increase of the structure factor peak $S(\pi,4n)$ from n to $n + 1$ as a function of $1/N$ with $N = 4n + 2$ using ED up to 24 spins, DMRG to 100 spins and linear extrapolation to the infinite chain; $S(\pi,g)$ diverges at $g = 0.20$, converges at $g = 0.25$. Inset: linear dependence of $S(\pi;g)$ on g near the origin for 16, 20 and 24 spins.

The π/2 peak for 8n spins reduces as expected to Eq. 6. In contrast to $S(\pi,g)$ at small g, however, there is no linear contribution in $1/g$ because $J_2 > 0$ is frustrating for either sign of $J_1$. The first-order correction $|\phi\rangle$ in $1/g$ is given by

$$(H_A + H_B - 2E_0)|\phi\rangle = -\frac{1}{g}\sum_p \vec{s}_p \cdot \vec{s}_{p+1}|G_A\rangle|G_B\rangle \quad (8)$$

$H_A$ and $H_B$ are HAFs on sublattices whose singlet GS and energy are $|G\rangle$ and $E_0$. Adjacent spins generate a singlet linear combination of triplets on each sublattice; $|\phi\rangle$ is a linear combination of such product states. Without explicitly solving Eq. 8, we obtain the general result for N

$$\langle\phi|\vec{s}_1 \cdot \vec{s}_{1+2p}|G_A\rangle|G_B\rangle = 0 \quad (9)$$

When both spins are in the same sublattice, the matrix element is zero since the triplet and GS of the other sublattice are orthogonal. It follows that $C(2p,g)$ and hence $S(\pi/2;g,8n)$ initially decreases as $1/g^2$.

Figure 4 shows incremental increases of $S(\pi/2;g,8n)$ from 8n to 8n + 8 spins as a function of $100/N$ with $N = 8n + 4$, followed by linear extrapolation to the infinite chain. The $1/g = 0$ points to 200 spins are $g = 0$ results to 100 spins. As noted above, shorter chains of 64 spins meet the requirement of $S(0;g) < 10^{-3}$ at large g. $S(\pi/2,g)$ converges and is clearly finite at $g = 2.0$ in the BOW phase. The estimated transition $g^{**}$ between the BOW and decoupled phases is around $1/g^{**} \sim 0.40$. As shown in the inset, $S(\pi/2;g,8n)$ initially goes as $-B_n/g^2$ with $B_6 = 0.17$ for 24 spins and is almost constant.

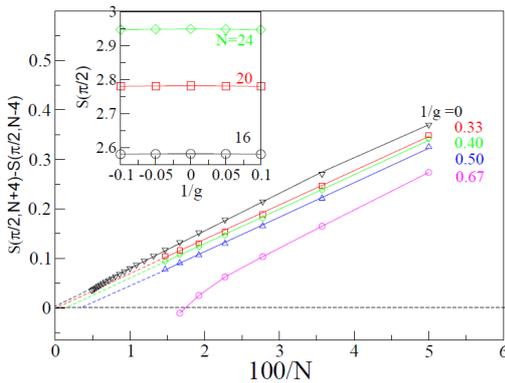

Fig. 4. Incremental increase of the structure factor peak $S(\pi/2,8n)$ from n to n + 1 as a function of $1/N$ with $N = 4n + 4$ using ED up to 24 spins, DMRG to 64 spins and linear extrapolation to the infinite chain; $S(\pi/2,g)$ diverges at $1/g = 0.33$, converges at $1/g = 0.50$. Inset: quadratic dependence of $S(\pi/2;g)$ on $1/g$ near the origin for 16, 20 and 24 spins; the maxima are at $1/g = 0$.

To summarize the $S(q_m;g)$ results, we return to Eq. 5. The divergences at $g = 0$ and $1/g = 0$ are identical. Since finite systems to 24 spins have $A_n > B_n$ by an order of magnitude, initial deviations $-B_n/g^2$ from $1/g = 0$ are much smaller than $-A_n g$ from $g = 0$. The BOW transition $g^*$ based on the divergence of $S(\pi;g)$ in Fig. 3 is consistent with $g^* = 0.2411$ based on $E_m = E_\sigma$. The transition at $1/g^{**} \sim 0.40$ from the BOW to decoupled phase in Fig. 4 is consistent with $g^{**} = 2.1$ based[14] on the $E_m = E_\sigma$ at large g. The $1/g^2$ decrease in Eq. 5 extends the decoupled phase to $J_1 < 0$.

## 4. Discussion

We have related the structure factor peak, $S(q_m;g)$, to the quantum phases of the $J_1J_2$ model, Eq. 1. $S(q_m;g)$ diverges at $q_m = \pi$ up to $g^*$ in the spin liquid phase with QLRO($\pi$), is finite in the frustrated BOW phase between $g^*$ and $g^{**}$, and diverges for $g > g^{**}$ in the gapless decoupled phase with QLRO($\pi/2$). We now address conflicting results that extend the BOW phase to $1/g = 0$.

To start with, theoretical and numerical works[1-13] have focused mainly on the quantum phase transition at $g^*$ to the BOW phase and recent studies[16-20] of Eq. 1 also deal with other sectors than large g. Interesting and exotic GS are generated by an external magnetic field, by anisotropic or antisymmetric rather than isotropic exchange, by changing the sign of $J_1$ or by increasing the range of exchange interactions. The magnetic properties of organic and inorganic crystals that contain spin chains provide other applications.

There are several reasons for a closer look at the $1/g \ll 1$ regime. First, the initial DMRG calculations[11] were limited to $1/g > 0.5$, far from the limit. Second, Okamoto and Namura[8] used ED in finite systems to obtain $g^*$ from the degeneracy $E_m = E_\sigma$; the same degeneracy at $1/g^{**}$ was not pointed out until later.[14] As a matter of consistency, ED in finite systems cannot decisive for locating the phase transition at $g^*$ but irrelevant at $g^{**}$. Third, exact HAF states describe both limits. ED of Eq. 1 with 4n spins yields n points $g_n$ with doubly degenerate GS and broken inversion symmetry, starting with $g_1 = 1/2$. The degenerate GS at the largest $g_n$ are closely related[21] to the product of sublattice ground states, $|G_A\rangle|G_B\rangle$, and the singlet linear combination of the lowest triplets, $^1|T_A\rangle|T_B\rangle$. In view of the insets to Figs. 3 and 4, it would be remarkable have a nondegenerate GS with divergent $S(\pi;g)$ up to $g^*$ while strictly limiting nondegenerate GS and divergent $S(\pi/2;g)$ to $1/g = 0$.

The large-g sector of Eq. 1 is particularly challenging, a point that may be relevant to spin chains as many-body problems. Field theories[11,13] starting with an HAF at g = 0 lead to different expressions for $E_m$ and rely on the same limited[11] DMRG for numerical support. Allen and Senechal[12] start with two HAFs at 1/g = 0 and discuss three different continuum descriptions of Eq. 1 along with various approximations. Turning to DMRG, we note that open boundary conditions (OBC) are typically used for an even number of spins. Quite aside from strong end effects,[15] inversion symmetry at sites is lost for even N. We find doubly degenerate GS and broken inversion symmetry at sites for N = 4n spins and PBC in Eq. 1 at n values of g. OBC not only lifts the degeneracy[21] but reverses the order to $E_m < E_\sigma$. Finally, accurate Monte Carlo methods have recently been devised[23] for 1D spin systems, including HAFs, but not for frustrated models due to a sign problem. Large g presents open questions for both field theory and numerical methods.

The magnitude of $S(q_m;g)$ bears directly on the quantum phases of frustrated spin chains. ED partly compensates for finite-size limitations by returning exact correlation functions $C(p,g)$. $S(q;g)$ is found directly for short-range correlation. Extrapolation to infinite chains is guided by the known HAF divergences of $S(\pi;0)$ or $S(\pi/2;\infty)$. But extrapolation entails approximations. Numerical methods and field theory are in agreement for the quantum transition of the $J_1 J_2$ model from the QLRO($\pi$) to BOW phase at g* = 0.2411, but disagree at present at large g. The peak $S(q_m;g)$ is finite in the BOW phase, diverges at $q_m = \pi$ for g < g* in the spin liquid phase with QLRO($\pi$) and at $q_m = \pi/2$ for 1/g > 1/g** in the decoupled phase. Frustrated spin chains whose GS has LRO($\pi$) at g = 0 undergo a first order quantum transition[21] with increasing g directly to the decoupled phase. The transition occurs at $g_c = 1/4\ln 2$ in an analytical model[21] with equal J = 2/(4n – 1) between spins in opposite sublattices and –J between spins in the same sublattice.

**Acknowledgments**: We thank D. Sen, A.W. Sandvik and S. Ramasesha for instructive discussions of BOW phase systems and the NSF for partial support of this work through the Princeton MRSEC (DMR-0819860). MK thanks DST for a Ramanujan Fellowship and support for thematic unit of excellence on computational material science.